\documentclass[reprint,notitlepage,twocolumn,
superscriptaddress,aps,longbibliography]{revtex4-1}
\usepackage[utf8]{inputenc}
\usepackage{braket}
\usepackage{amsmath}
\usepackage{amsthm}
\usepackage{mathrsfs}
\usepackage{mathtools}
\usepackage{amssymb}
\usepackage{dsfont}
\usepackage{braket}
\usepackage{bm}
\DeclareMathOperator{\tr}{tr}
\usepackage{color}
\pdfoutput=1

\usepackage[english]{babel}
\usepackage[T1]{fontenc}
\usepackage{amsmath}
\usepackage[colorlinks=true]{hyperref}
\usepackage{enumitem}
\usepackage{amsthm}
\usepackage{amssymb}
\usepackage{graphicx}
\usepackage{float}
\usepackage[ruled]{algorithm2e}

\begin{document}

\title{
 Learning quantum properties from short-range correlations using multi-task  networks}

\author{Ya-Dong Wu}
\affiliation{John Hopcroft Center for Computer Science, Shanghai Jiao Tong University, Shanghai 200240, China}
\affiliation{QICI Quantum Information and Computation Initiative, Department of Computer Science,
The University of Hong Kong, Pokfulam Road, Hong Kong}
\author{Yan Zhu}
\email{yzhu2@cs.hku.hk}
\thanks{Ya-Dong Wu and Yan Zhu contribute equally}
\affiliation{QICI Quantum Information and Computation Initiative, Department of Computer Science,
The University of Hong Kong, Pokfulam Road, Hong Kong}
\author{Yuexuan Wang}
\affiliation{AI Technology Lab, Department of Computer Science,
The University of Hong Kong, Pokfulam Road, Hong Kong}
\affiliation{ College of Computer Science and Technology,
Zhejiang University, Zhejiang Province, China}
\author{Giulio Chiribella}
\email{giulio@cs.hku.hk}
\affiliation{QICI Quantum Information and Computation Initiative, Department of Computer Science,
The University of Hong Kong, Pokfulam Road, Hong Kong}
\affiliation{Department of Computer Science, Parks Road, Oxford, OX1 3QD, United Kingdom}
\affiliation{Perimeter Institute for Theoretical Physics, Waterloo, Ontario N2L 2Y5, Canada}

\begin{abstract}
 Characterizing  multipartite   quantum systems is  crucial  for quantum computing and  many-body physics.  The problem, however, becomes  challenging when the  system size is large and  the properties of interest involve correlations among  a large number of particles. 
Here we introduce  a neural network model that can predict various quantum properties of many-body quantum states  with constant correlation length, using only measurement data from a small number of neighboring sites. The model is based  on the technique of multi-task learning, which we show to offer several advantages over traditional single-task approaches.  
Through numerical experiments, we show that multi-task learning can be applied  to sufficiently regular states to predict global properties, like string order parameters, from the observation of short-range correlations, and to distinguish  between quantum phases that cannot be distinguished by single-task networks.
  Remarkably, our  model appears to be able to   transfer information learnt from lower dimensional quantum systems to higher dimensional ones, and to make accurate predictions for  Hamiltonians that were not seen in the training. 
\end{abstract}

\maketitle

\section{Introduction}

The experimental characterization of many-body quantum states   is an essential task in quantum information and  computation. Neural networks provide a powerful approach to quantum state characterization    
~\cite{torlai2018,carrasquilla2019,zhu2022,schmale2022efficient}, enabling a compact representation of sufficiently structured quantum states~\cite{carleo2017}. 
In recent years, different types of neural networks  have been successfully utilized to predict  properties of quantum systems, including quantum fidelity~\cite{zhang2021,xiao2022, du2023shadownet} and other measures of  similarity~\cite{wu2023, qian2023multimodal}, quantum entanglement~\cite{gao2018,gray2018,koutny2023deep}, entanglement entropy~\cite{torlai2018,torlai2019,huang2022measuring}, two-point correlations~\cite{torlai2018,torlai2019,carrasquilla2019,kurmapu2022reconstructing} and Pauli expectation values~\cite{smith2021,schmale2022efficient}, as well as to identify phases of matter~\cite{carrasquilla2017machine,van2017,huembeli2018,rem2019,kottmann2020}.  

A challenge in characterizing multiparticle quantum systems is that important properties,  such as topological invariants characterizing different quantum phases of matter~\cite{PhysRevB.86.125441}, are global:  their direct estimation requires  measurements that probe the  correlations among a large number of particles.   For example,  randomized measurement techniques~\cite{huang2020,huang2022learning,elben2023randomized, zhao2023learning} provide an effective way to characterize global properties from  measurements performed locally on individual particles, but generally require information about the correlations among a large number of local measurements.  
Since estimating multiparticle correlations    becomes difficult when the system size scales up,  it would be desirable to  have a way to learn global properties from  data collected only from  a small number of neighboring sites.
So far, the characterization of many-body quantum states from short-range correlations has been investigated for the purpose of quantum state tomography~\cite{cramer2010efficient,baumgratz2013,lanyon2017,guo2023}.
For large quantum systems, however,  a  full tomography becomes challenging, and   efficient methods for learning quantum properties from short-range correlations are still missing.


 In this paper, we develop  a neural network model that can predict various quantum properties of many-body quantum states  from short-range correlations.  
Our model utilizes the technique of 
multi-task learning~\cite{zhang2021survey} to generate concise state representations that integrate diverse types of information. In particular, the model can integrate information obtained from few-body measurements into a representation of the overall quantum state, in a way that is reminiscent of the quantum marginal problem \cite{klyachko2006quantum,christandl2006spectra,schilling2015quantum}. 
 The state representations produced by our model are  then  used to  learn new physical properties that were not seen  during the training, including global properties such as string order parameters and many-body topological invariants~\cite{PhysRevB.86.125441}.

 For ground states with short-range correlations, we find that our model accurately predicts  nonlocal features using only   measurements on a few nearby particles.  With respect to traditional, single-task neural  networks, our model achieves more precise predictions with comparable amounts of input data, and enables a direct unsupervised classification of symmetry protected topological (SPT) phases that could not be distinguished in the single-task approach. In addition, we find that, after the training is completed, the model can be applied to quantum states and  Hamiltonians outside the original training set, and even to quantum systems of higher dimension.  This strong performance on out-of-distribution states suggests that our multi-task network could be used  as a tool to explore the next frontier of intermediate-scale quantum systems.

\section{Results}
\label{sec:rsults}

\begin{figure*}
    \centering
    \includegraphics[width=0.98\textwidth]{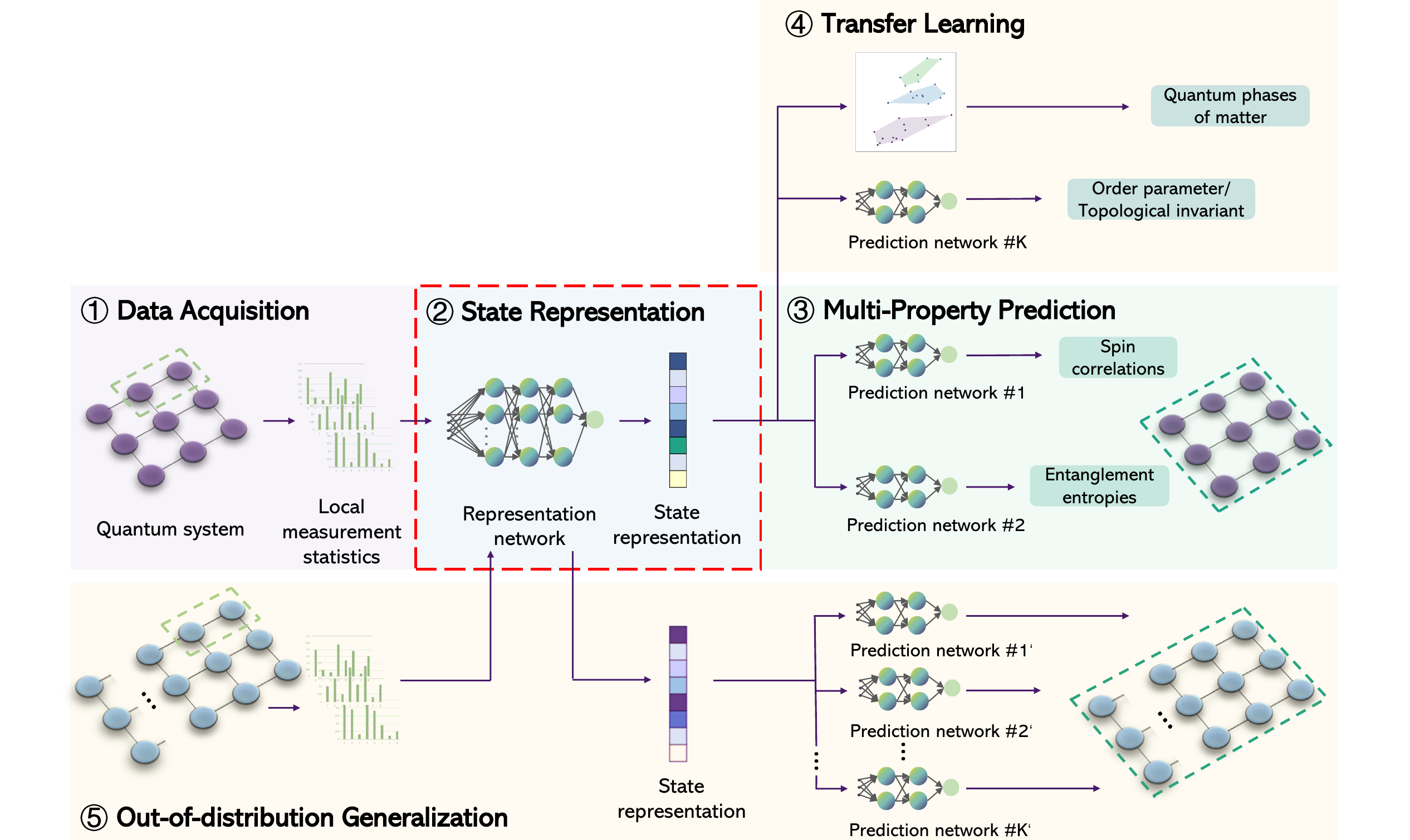}
    \caption{Flowchart of our multi-task neural network.    
  In the data acquisition phase (1), the experimenter performs short-range local measurements on the system of interest.  The resulting data is used  to produce a concise representation of the quantum state (2). The state representation is then fed into a set of prediction networks, each of which generates predictions for a given type of quantum property (3). After the state representation network and prediction networks are jointly trained,  the state representations are employed in new tasks, such as  unsupervised classification of quantum phases of matter, or prediction of order parameters and topological invariants (4). Once trained, the overall model can generally be applied to  out-of-distribution quantum states and higher-dimensional quantum systems (5).  }
    \label{fig1}
\end{figure*}

\subsection*{Multi-task framework for quantum properties.}  
  Consider the scenario where an experimenter has access to multiple copies of an unknown quantum state $\rho_{\bm{\theta}}$, characterized by some physical parameters $\bm{\theta}$. For example, $\rho_{\bm \theta}$ could be a ground state of  many-body local Hamiltonian depending on  $\bm \theta$.  The experimenter's goal is  to predict a set of properties of the quantum state,  such as the expectation values of some observables, or some  nonlinear functions, such as the von Neumann entropy. 
The experimenter is able to perform a restricted set of quantum measurements, denoted by $\mathcal M$.  Each measurement $\bm M  \in  \cal M$ is described by a positive operator-valued measure (POVM)  $\bm M  =  (M_j)$, where the index $j$ labels the measurement outcome,  each $M_j$ is a positive operator acting on the system's Hilbert space, and the normalization condition $\sum_j  M_j  =  I$ is satisfied.  
 In general, the measurement set $\mathcal{M}$ may not  be  informationally complete.  For multipartite systems, we will typically take $\mathcal M$ to consist of local measurements performed on a small number of  neighboring systems. 

To collect data, the experimenter randomly picks a subset of measurements  $\mathcal{S} \subset  \cal M$,  and performs them on different copies of the state  $\rho_{\bm{\theta}}$.  We will denote by $s$ the number of measurements in $\cal S$, and  by  $\bm{M}_i:=\left( M_{ij}\right)$ the $i$-th POVM in $\cal S$.  
For simplicity, if not specified otherwise,  we assume that each measurement in $\cal S$ is repeated sufficiently many  times that the experimenter can reliably estimate the outcome distribution $\bm{d}_i:=\left(d_{ij}\right)$, where $d_{ij}:=\tr\left(\rho M_{ij}\right)$.

The experimenter's goal is to predict multiple quantum properties of $\rho_{\bm{\theta}}$ using the outcome  distributions $(\bm{d}_i)_{i=1}^s$. 
  This task is achieved by a  neural network  that consists of an encoder and multiple decoders, where the encoder $\mathcal E$  produces a representation of quantum states and the $k$-th decoder $\mathcal D_k$ produces a prediction of the $k$-th property of interest.  Due to their roles, the encoder and decoders are also known as representation and prediction networks, respectively.

The input of the representation network  $\cal E$ is  the outcome distribution $\bm{d}_i$,   together with a parametrization of the corresponding measurement $\bm{M}_i$, hereafter denoted by $\bm{m}_i$.   From the pair of data $(\bm{d}_i,  \bm{m}_i)$, the network  produces a state representation $\bm{r}_i:=\mathcal E(\bm{d}_i, \bm{m}_i)$.   To combine the state representations arising from different measurements in $\cal S$, the network computes the average  $\bm{r}:=\frac{1}{s}\sum_{i=1}^s \bm{r}_i$. At this point, the vector $\bm{r}$ can be viewed as a representation of the unknown quantum state $\rho$. 

Each prediction network $\mathcal{D}_k$ is dedicated to a different property of the quantum state.     In the case of multipartite quantum systems, we include the option of evaluating the property on a subsystem, specified by a parameter  $q$.   We denote by    $f_{k,q}(\rho_{\bm{\theta}})$ the correct value of the $k$-th property of subsystem $q$ when the total system is in the state $\rho_{\bm \theta}$.  
 Upon receiving the state representation $\bm r$ and the subsystem specification $q$, the prediction network produces an estimate $\mathcal D_k(\bm{r}, q)$ of the value  $f_{k,q}(\rho)$.

The representation network and all the prediction networks are trained jointly,  with the goal of minimizing the prediction error on a set of fiducial states.  The fiducial states are chosen by  randomly sampling a set of physical parameters  $(\bm{\theta}_l)_{l=1}^L$.  For each fiducial state $\rho_{\bm{\theta_l}}$,  
we independently sample a set  of measurements $\mathcal S_l$  and calculate the outcome distributions for each measurement in the set $\mathcal S_l$.  We randomly choose a subset of properties $\mathcal K_l$ for each $\rho_{\bm{\theta_l}}$, where each property $k\in  \mathcal K_l$ corresponds to a set of subsystems $\mathcal Q_k$, 
and then calculate the correct values of the quantum properties $\{f_{k,q}(\rho_{\bm{\theta}_l})\}$ for all properties $k\in  \mathcal K_l$ associated with subsystems $q\in  \mathcal Q_k$. 
The training data may be either classically simulated or gathered by actual measurements on the set of fiducial states, or it could also be obtained by any combination of these two approaches.

During the training,  we do not provide the model with any information about the physical parameters $\bm{\theta}_l$ or about the functions $f_{k,q}$.  
   Instead, the  internal parameters  of the neural networks are jointly  optimized in order to  minimize the estimation errors $|\mathcal D_k\left( 1/s\sum_{i=1}^s \mathcal E(\{\bm{d}_i, \bm{m}_i\}), q\right)- f_{k, q}(\rho_{\bm{\theta}})|$ summed over all the fiducial states,   all chosen  properties,  
   and all chosen subsystems.

After the training is concluded, our  model can be used for predicting quantum properties, either within the set of properties seen during training or outside this set.  The requested properties are predicted on a new,  unknown state $\rho_{\bm{\theta}}$, and even out-of-distribution state $\rho$ that has a structural similarity with the states in the original distribution, e.g., a ground state of the same type of  Hamiltonian, but for a quantum system with a larger number of particles. 

The high-level structure of our model is illustrated in Figure~\ref{fig1}, while the details of the neural networks are presented in Methods. 



\subsection*{Learning ground states of Cluster-Ising model}
We first test the performance of our model on a relatively small system of  $N=9$ qubits whose  properties can be explicitly calculated.    For the state family, we take the  ground states of one-dimensional cluster-Ising model~\cite{smacchia2011}
\begin{equation}\label{eq:clusterIsing}
    H_{\text{cI}}=-\sum_{i=1}^{N-2}\sigma_i^z \sigma_{i+1}^x \sigma_{i+2}^z-h_1\sum_{i=1}^N \sigma_i^x -h_2\sum_{i=1}^{N-1}\sigma_i^x \sigma_{i+1}^x.
\end{equation}
The  ground state falls in one of  three phases, depending on the values of the parameters $(h_1, h_2)$. The three phases  are: the SPT phase, 
the paramagnetic phase, and  the antiferromagnetic phase.  SPT phase can be distinguished from two other phases by measuring the string order parameter~\cite{cong2019quantum,herrmann2022}
$\braket{\tilde{S}}:=\braket{\sigma_1^z \sigma_2^x \sigma_4^x \dots \sigma_{N-3}^x \sigma_{N-1}^x \sigma_N^z}$, which is a global property involving $(N+3)/2$ qubits.

We test our network model  on the ground states corresponding to   a $64\times 64$ square grid in the parameter region  $(h_1, h_2)\in [0, 1.6]\times [-1.6, 1.6]$.
For the  set of accessible measurements $\mathcal M$, we take  all possible three-nearest-neighbour Pauli  measurements, corresponding to the observables   $\sigma_i^\alpha \otimes\sigma_{i+1}^\beta \otimes\sigma_{i+2}^\gamma$, where $i\in\{1,2,\dots, N-2\}$ and $\alpha,\beta,\gamma\in \{x,y,z\}$.

\begin{figure*}
    \centering
   \includegraphics[width=0.9\textwidth]{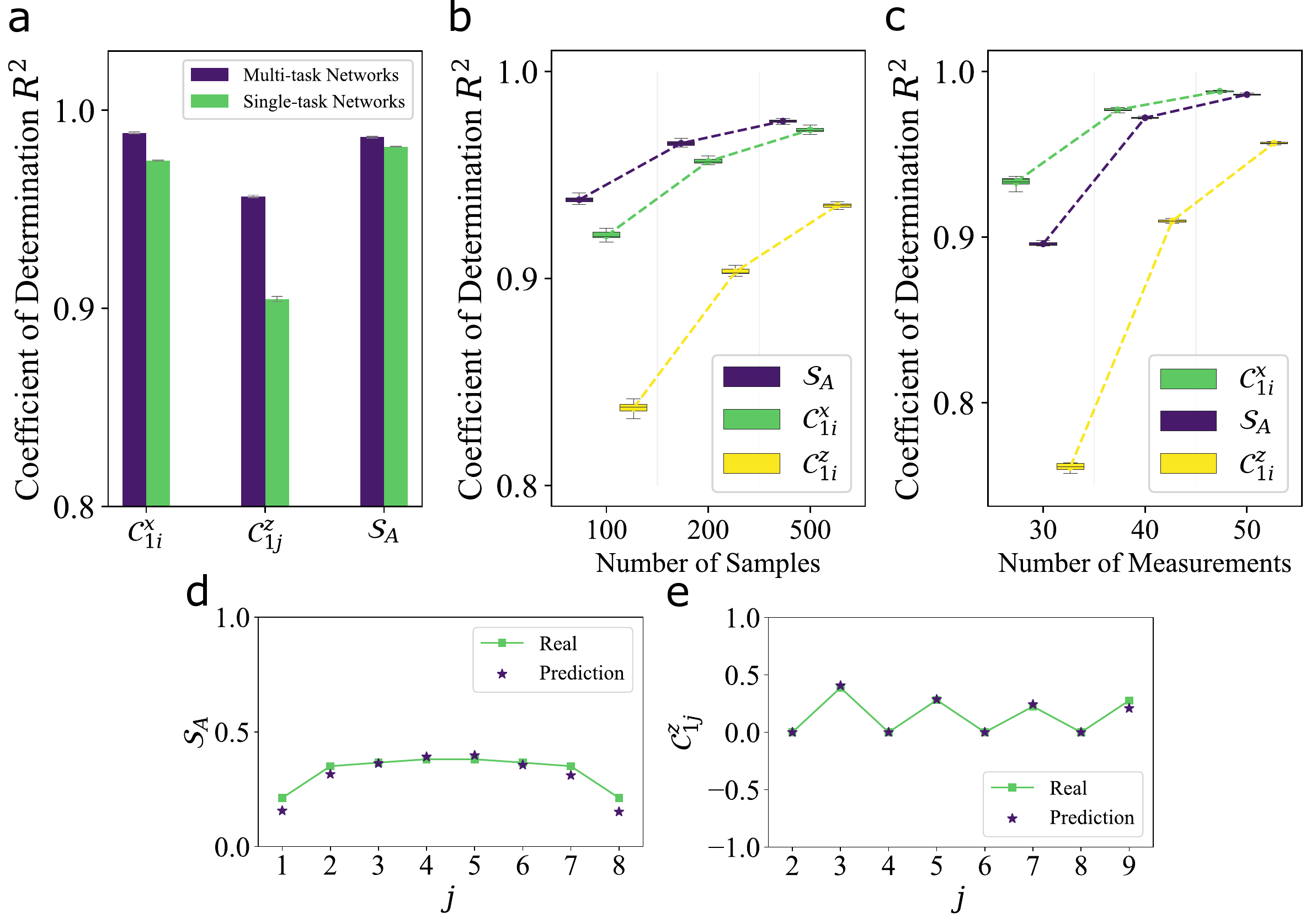}
    \caption{Predicting  properties of ground states of cluster-Ising model. Subfigure {\bf a} compares the prediction accuracy between our multi-task model and single-task models for predicting  two-point correlation functions $\mathcal C_{1j}^x$ and $\mathcal C_{1j}^z$, and entanglement entropy $\mathcal S_A$.
    Subfigures {\bf b} and {\bf c} show how the number of samples for each measurement and the number of measurements affect the coefficient of determination for the predictions of $\mathcal S_A$, $\mathcal C_{1j}^x$ and $\mathcal C_{1j}^z$, respectively.
    Subfigures {\bf d} and {\bf e} show the predictions of $S_A$ and $\mathcal C_{1j}^z$ for a ground state near phase transition marked by a red star in Subfigure \ref{fig3}{\bf d}.
}
    \label{fig2}
\end{figure*}

For the prediction tasks, we consider two properties:  (A1)  the two-point correlation function $\mathcal C_{1j}^\alpha:=\braket{\sigma_1^\alpha \sigma_{j}^\alpha}_\rho$, where $1< j \le N$ and $\alpha=x, z$; (A2) the  R\'{e}nyi entanglement entropy of order two $ S_A:=-\log_2 \left(\tr\rho_A^2\right)$ for subsystem $A=[1,2,\dots, i]$, where $1\le i< N$.
Both properties (A1) and (A2) can be either numerically evaluated,  or experimentally estimated by preparing the appropriate quantum state  and performing  randomized measurements~\cite{elben2023randomized}.

We train our neural network with respect to the fiducial ground states corresponding to $300$ randomly chosen points from our 4096-element grid.  For each fiducial state, we provide the neural network with the outcome distributions of  $s=50$ measurements,  randomly chosen from  the 243 measurements in $\mathcal M$. Half of these fiducial states randomly chosen from the whole set are labeled by the values of property (A1) and the other half are labeled by property (A2). 
After training is concluded, we apply our trained model to predicting  properties (A1)-(A2) for all remaining  ground states corresponding to points on the  grid. For each test state, the representation network  is provided with the outcome  distributions
on $s=50$ measurement settings randomly chosen from $\mathcal M$. 

Figure~\ref{fig2}{\bf a} illustrates the coefficient of determination ($R^2$), averaged over all test states,  for each type of property.  Notably, all the values of $R^2$ observed in our experiments are above $0.95$. 
Our network makes accurate predictions even near the boundary between the  SPT phase and paramagnetic phase, in spite of the fact that phase transitions typically  make it  more difficult to capture the ground state properties from limited measurement data.  For a ground state close to the boundary,  marked by a star in the phase diagram (Figure~\ref{fig3}{\bf d}), the predictions of the entanglement entropy $\mathcal S_{A}$ and spin correlation $\mathcal C_{1j}^z$ are close to  the corresponding ground truths, as shown  in Figures~\ref{fig2}{\bf d} and~\ref{fig2}{\bf e}, respectively.

In general,  the accuracy of the predictions depends on the number of samplings for each measurement as well as the number of measurement settings. For our experiments, the dependence is illustrated in Figures~\ref{fig2}{\bf b} and~\ref{fig2}{\bf c}.

To examine whether our multi-task neural network model enhances the prediction accuracy compared to single-task networks,  we perform  ablation experiments~\cite{cohen1988evaluation}. We train three individual single-task neural networks as our baseline models, each of which predicts spin correlations in Pauli-x axis, spin correlations in Pauli-z axis, and entanglement entropies, respectively.  For each single-task neural network, the training provides the network with the corresponding properties for the $300$ fiducial ground states, without providing any information  about the other properties. After the training is concluded, we apply each single-task neural network to predict the corresponding properties on all the test states and use their predictions as baselines to benchmark the performance of our multi-task neural network. Figure~\ref{fig2}{\bf a} compares the values of $R^2$ for the predictions of our multi-task neural model with those of the single-task counterparts. The results demonstrate that learning multiple physical properties simultaneously enhances the prediction of each individual  property.

\subsection*{Transfer learning  to new tasks}

\begin{figure*}
    \centering
    \includegraphics[width=0.9\textwidth]{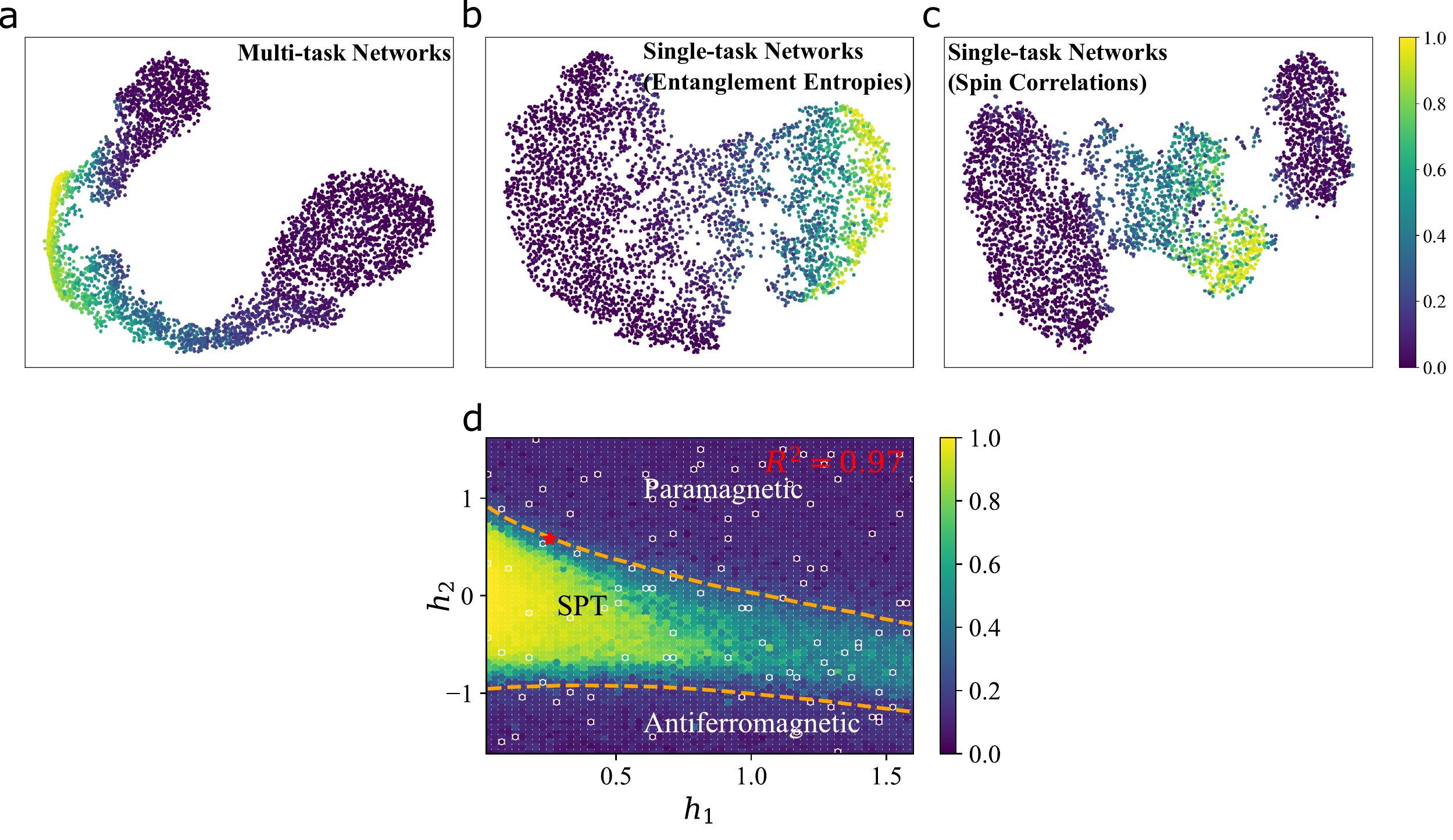}
    \caption{Transfer learning to predict properties of the ground states of the cluster-Ising model. Subfigures {\bf a, b} and {\bf c} illustrate the 2D projection of the state representations obtained with the t-SNE algorithm, where the color of each data point indicates the true value of the string order parameter $\braket{\tilde{S}}$ of the corresponding ground state. Subfigure {\bf a} corresponds to the state representations produced for jointly predicting  spin correlations and entanglement entropy. Subfigures {\bf b} and {\bf c} correspond to the state representations produced for separately predicting entanglement entropy and spin correlations, respectively. Subfigure {\bf d} shows the predictions of $\braket{\tilde{S}}$ for the ground states corresponding to a $64\times 64$ grid in  parameter space, together with the true values of $\braket{\tilde{S}}$ for $100$  randomly chosen states indicated by white circles, where the dashed curves are the phase boundaries between SPT phase and the other two phases.}
    \label{fig3}
\end{figure*}

We now show that  the state representations produced by the encoder can  be used to perform new tasks that were  not encountered during the training phase.
In particular, we show that state representations can be used to distinguish between the phases of matter associated to different values of the Hamiltonian parameters in an unsupervised manner.    To this purpose,    we project the representations of all the test states onto a two-dimensional (2D) plane using the t-distributed stochastic neighbour embedding (t-SNE) algorithm. 

The results are shown in Figure~\ref{fig3}{\bf a}.
Every data point shows  the exact value of the  string order parameter, which distinguishes between the SPT phase and the other two phases.   
Quite strikingly, we find that the disposition of the points in the 2D representation matches the values of the string order parameter,  even though no information about the string order parameters was provided during the training, and even though the string order is a global property, while the measurement data provided to the network came from a small number of neighboring sites.

A natural question is whether the accurate classification of phases of matter observed above is a consequence of the multi-task nature of our model.  To shed light into this question, we compare the results of our multi-task  network  with those of  single-task neural networks, feeding the state representations generated by these networks into the t-SNE algorithm to produce a 2D representation. 
The pattern of the projected state representations in Figure~\ref{fig3}{\bf b} indicates that when trained only with the values of entanglement entropies, the neural network cannot distinguish between the paramagnetic phase and the antiferromagnetic phase.
Interestingly, a single-task network trained only on the spin correlations can still distinguish the SPT phase from the other two phases, as shown in Figure~\ref{fig3}{\bf c} .   However, in the next section we see  that applying  random local gates induces errors in the single-task network, while the multi-task network still achieves a correct classification of the different phases. 


Quantitatively, the values of the string order parameter can be extracted from the state representations using another neural network  $\mathcal N$.   To train this network,   we randomly pick  $100$ reference states $\{\sigma_i\}$ out of the $300$ fiducial states  and minimize the error $\sum_{i=1}^{100} |\mathcal N(\bm{r}_{\sigma_i})-\braket{\tilde{S}}_{\sigma_i}|$. Then, we use the trained   neural network $\mathcal N$ to produce the prediction $\mathcal N(\bm{r}_\rho)$  of $\braket{\tilde{S}}_\rho$ for every other state $\rho$. The prediction for each ground state is shown in the phase diagram (Figure~\ref{fig3}{\bf d}), where the $100$ reference states are marked by white circles. The predictions are close to the true values of string order parameters in Figure~\ref{fig5}{\bf c}.    It is  important to stress that, while the network $\cal N$ was trained on values of the string order parameter, the representation network $\cal E$ was not provided any  information about this parameter.     Note also that the values of the Hamiltonian parameters $(h_1, h_2)$ are just provided in the figure for the purpose of visualization: in fact, no information about the Hamiltonian parameters was provided to  the  network during training or test.  
  In Supplementary Note 6, we show that our neural network model trained for predicting entanglement entropy and spin correlations can also be transferred to other ground-state  properties of the cluster-Ising model.

\subsection*{Generalization to out-of-distribution states}
In the previous sections, we  assumed that  both the training and the testing states were randomly sampled from a set of ground states of the cluster-Ising model~(\ref{eq:clusterIsing}). 
In this subsection, we explore how a model trained on a given set of quantum states can  generalize to states outside the original set in an unsupervised or weakly supervised manner.

 Our first finding is that  our model, trained on the ground states of the cluster-Ising model, can effectively cluster general quantum states in the SPT phase and the trivial phase (respecting the symmetry of bit flips at even/odd sites), without further training. Random quantum states in SPT (trivial) phase can be prepared by applying short-range  symmetry-respecting local random quantum gates on a cluster state in the SPT phase (a product state $\ket{+}^{\otimes N}$ in the paramagnetic phase). For these  random quantum states, we follow the same  measurement strategy adopted before, feed the measurement data into our trained representation network, and use t-SNE to project the state representations onto a 2D plane. 

\begin{figure*}
    \centering
    \includegraphics[width=0.6\textwidth]{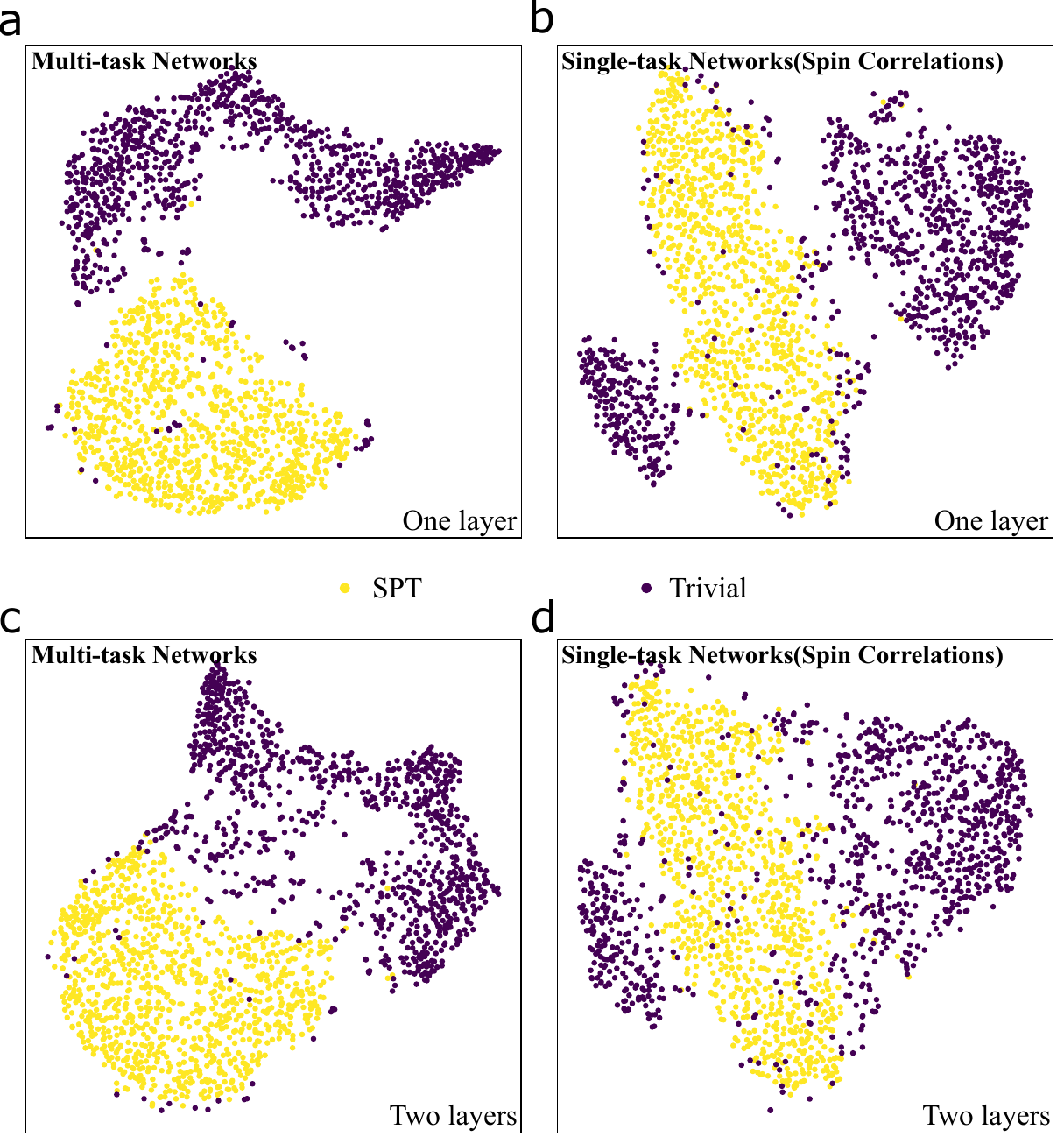}
    \caption{2D projections of state representations for those states prepared by shallow random symmetric quantum circuits. 
    Subfigures {\bf a} and {\bf b} correspond quantum states in the SPT and the trivial phases prepared by one layer of random quantum gates, and Subfigures {\bf c} and {\bf d} correspond quantum states in the SPT and the trivial phases prepared by two layers of random quantum gates. 
    Subfigures {\bf a} and {\bf c} illustrate state representations produced by our multi-task neural network. Subfigures {\bf b} and {\bf d} illustrate state representations produced by the neural network trained only on spin correlations. }
    \label{fig4}
\end{figure*}

When the quantum circuit consists of a layer of translation-invariant next-nearest neighbour symmetry-respecting random gates, our model successfully classifies the output states into their respective SPT phase and trivial phase in both cases, as shown by Figure~\ref{fig4}{\bf a}. In contrast, feeding the same measurement data into the representation network trained only on spin correlations fails to produce two distinct clusters via t-SNE, as shown by Figure~\ref{fig4}{\bf b}. While this neural network successfully classifies different phases for the cluster-Ising model, random local quantum gates confuse it. This failure is consistent with 
the recent observation that extracting linear functions of a quantum state is insufficient for classifying arbitrary states within SPT phase and trivial phase~\cite{huang2022provably}.

We then prepare more complex states by applying two layers of translation-invariant random gates consisting of both nearest neighbour and next-nearest neighbour gates preserving the symmetry onto the initial states. The results in Figure~\ref{fig4}{\bf c} show that the state representations of these two phases remain different, but the boundary between them in the representation space is less clearly identified. Whereas, the neural network trained only on spin correlations fails to classify these two phases, as shown by Figure~\ref{fig4}{\bf d}.

\begin{figure*}
    \centering
    \includegraphics[width=0.8\textwidth]{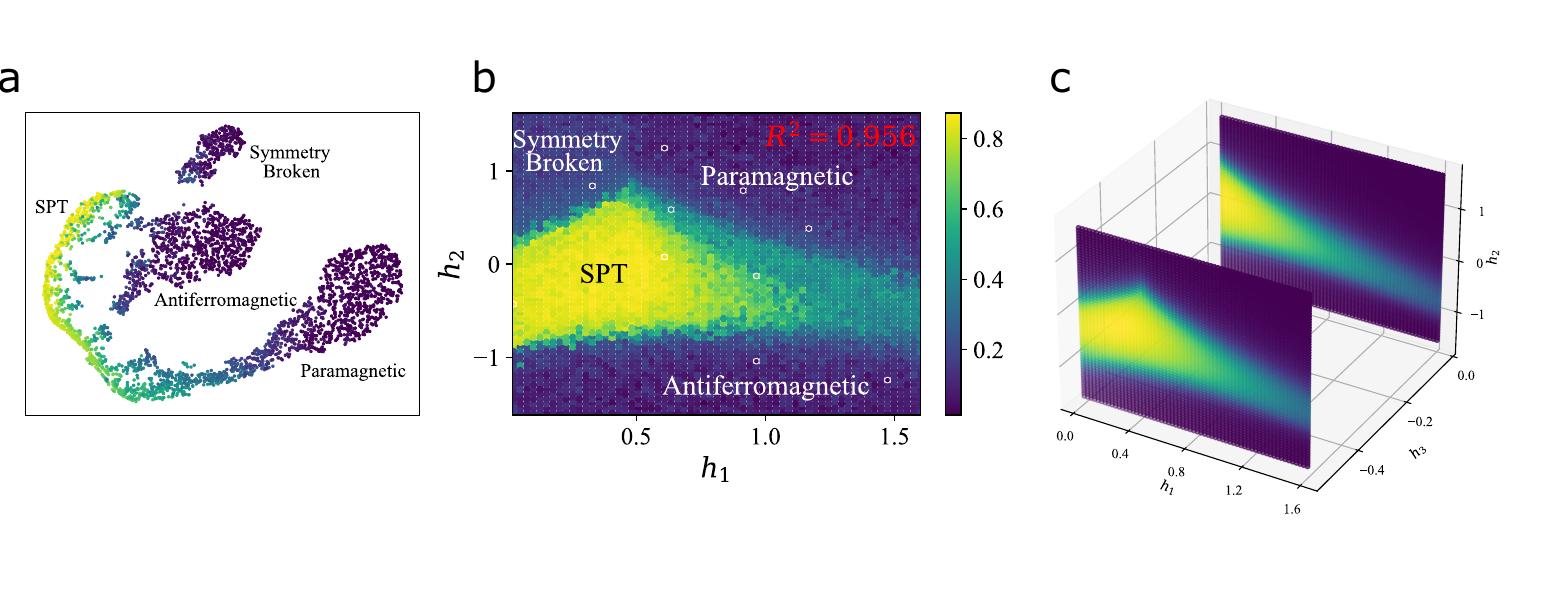}
    \caption{ Prediction of properties of ground states of a perturbed Hamiltonian. Subfigure {\bf a} illustrates the 2D projections of state representations for the ground states of the perturbed Hamiltonian, together with their true values of $\braket{\tilde{S}}$.  Subfigure {\bf b} illustrates the predictions of $\braket{\tilde{S}}$ using our adjusted neural network for the perturbed model. Subfigure {\bf c} show the true values of string order parameters $\braket{\tilde{S}}$ for both the original model~(\ref{eq:clusterIsing}) and the perturbed model~(\ref{eq:perturb}). }
    \label{fig5}
\end{figure*}

Finally, we demonstrate that our neural model, trained on the cluster-Ising model, can adapt to learn the ground states of a new, perturbed Hamiltonian~\cite{liu2023}
\begin{equation}\label{eq:perturb}
    H_{\text{pcI}}= H_{\text{cI}} +h_3 \sum_{i=1}^{N-1}\sigma_i^z \sigma_{i+1}^z.
\end{equation} 
This perturbation breaks the original symmetry, shifts the boundary of the cluster phase, and introduces a new phase of matter.  In spite of these substantial changes,  Figure~\ref{fig5}{\bf a} shows that our model, trained on the unperturbed cluster-Ising model, successfully identifies the different phases, including the new phase from the perturbation.  Moreover, using just $10$ randomly chosen additional reference states (marked by white circles in Figure~\ref{fig5}{\bf b}),  the original prediction network can be adjusted to predict the values of $\braket{\tilde{\mathcal S}}$ from state representations. As shown in Figure~\ref{fig5}{\bf b}, the predicted values closely match the ground truths in Figure~\ref{fig5}{\bf c}, achieving a coefficient of determination of up to $0.956$ between the predictions and the ground truths.

\subsection*{Learning ground states of XXZ model}
We now apply our model to a larger quantum system, consisting of 50 qubits in ground states of  the bond-alternating XXZ model~\cite{elben2020many} 
\begin{align}\notag   H=&J\sum_{i=1}^{N/2}\left(\sigma_{2i-1}^x \sigma_{2i}^x +\sigma_{2i-1}^y \sigma_{2i}^y+\delta\sigma_{2i-1}^z \sigma_{2i}^z\right)  \\    &+  J'\sum_{i=1}^{N/2-1}\left(\sigma_{2i}^x \sigma_{2i+1}^x +\sigma_{2i}^y \sigma_{2i+1}^y+\delta \sigma_{2i}^z \sigma_{2i+1}^z\right) \, , \label{eq:XXZmodel}
\end{align}
where  $J$ and $J'$ are the alternating values of the  nearest-neighbor spin couplings.   We consider a set of ground states corresponding to a $21\times 21$ square grid in the parameter region $(J/J', \delta)\in (0, 3)\times (0,4)$.
Depending on the ratio of $J/J'$ and the strength of $\delta$, the corresponding ground state falls into one of three possible phases: trivial SPT phase, topological SPT phase, and symmetry broken phase.  

Unlike   the SPT phases of cluster-Ising model, the SPT phases of bond-alternating XXZ model cannot be detected by any string order parameter.  Both SPT phases  
are protected by bond-center inversion symmetry, and  detecting them requires  a many-body topological invariant, called the partial reflection topological invariant~\cite{elben2020many} and denoted by
\begin{equation}
\mathcal Z_{\text{R}}:=\frac{\tr(\rho_I \mathcal R_I)}{\sqrt{\left[\tr(\rho_{I_1}^2)+\tr(\rho_{I_2}^2)\right]/2}} \, .
\end{equation}
Here, 
$\mathcal R_I$ is the swap operation on subsystem $I:=I_1\cup I_2$ with respect to the center of the spin chain, and $I_1=[N/2-5, N/2-4, \dots, N/2]$ and $I_2=[N/2+1, N/2+2, \dots, N/2+6]$ are two subsystems with six qubits.

\begin{figure*}
    \centering
    \includegraphics[width=0.9\textwidth]{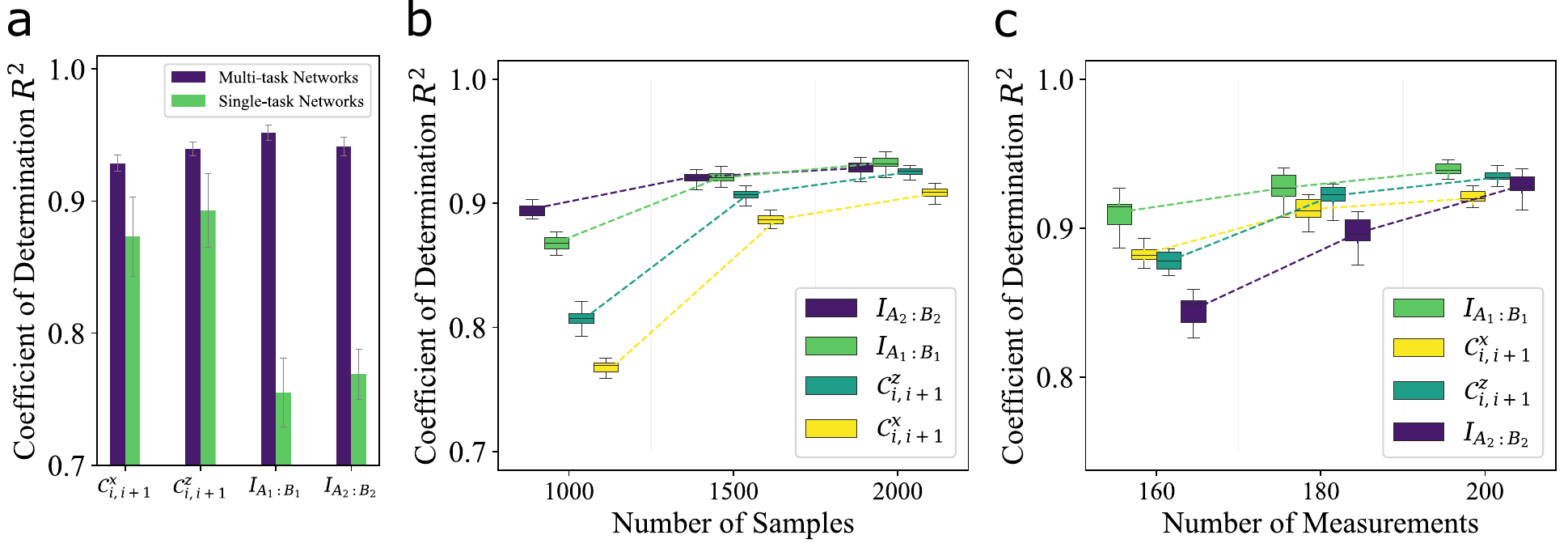}
    \caption{
    Predicting  properties of $50$ qubit ground states of bond-alternating XXZ model. Subfigure {\bf a} compares the prediction accuracy between our multi-task model and single-task models for predicting  spin correlations $\mathcal C_{i\, i+1}^x$ and $\mathcal C_{i\, i+1}^z$, as well as R\'{e}nyi mutual information $I_{A:B}$. Subfigure {\bf b} and {\bf c} show how the number of samples for each measurement and the number of measurements affect the coefficient of determination for the predictions of all the properties.}
    \label{fig6}
\end{figure*}

 For the set of  possible measurements $\mathcal M$, we take all  possible three-nearest-neighbour Pauli projective measurements, as we did earlier in the cluster-Ising model. 
 For the prediction tasks, we consider  two types of quantum properties: (B1) nearest-neighbour spin correlations $\mathcal C_{i,i+1}^\beta:=\braket{\sigma_i^\beta\sigma_{i+1}^\beta} (1\le i\le N-1)$, where $\beta=x, z$; (B2)  order-two R\'{e}nyi mutual information  $I_{A:B}$, where $A$ and $B$ are both $4$-qubit subsystems: either $A_1=[22:25], B_1=[26:29]$ or $A_2=[21:24], B_2=[25:28]$.

We train our neural network with respect to the fiducial ground states corresponding to $80$ pairs of $(J/J', \delta)$,  randomly sampled from the 441-element grid. For each fiducial state, we provide the neural network with the  probability distributions corresponding to $s=200$ measurements randomly chosen from the 1350 measurements in  $\mathcal M$.
Half of the fiducial states randomly chosen from the entire set are labeled by the property of (B1), while the other half are labeled by the property of (B2).
After the training is concluded, we use our trained model to predict both properties (B1) and (B2) for all the ground states in the grid.

Figure~\ref{fig6}{\bf a} demonstrates the strong predictive performance of our model, where the values of $R^2$ are above $0.92$ for all properties averaged over test states.
We benchmark the performances of our multi-task neural network with the predictions of single-task counterparts. Here each single-task neural network, the size of which is same as the multi-task network, aims at predicting one single physical property and is trained using the same set of measurement data of $80$ fiducial states together with one of their properties: $C_{i,i+1}^x$, $C_{i,i+1}^z$, $I_{A_1:B_1}$ and $I_{A_2:B_2}$. Figure~\ref{fig6}{\bf a} compares the coefficients of determination for the predictions of both our multi-task neural network and the single-task neural networks, where each experiment is repeated multiple times over different sets of $s=200$ measurements randomly chosen from $\mathcal M$. The results indicate that  our multi-task neural model not only achieves higher accuracy in the predictions of all properties, but also is much more robust to different choices of quantum measurements. 
As in the case of the cluster-Ising model, we  also study how the number of quantum measurements $s$ and the number of samplings for each quantum measurement affect the prediction accuracy of our neural network model, as shown by Figures~\ref{fig6}{\bf b} and~\ref{fig6}{\bf c}. Additionally, we test how the size of the quantum system affects the prediction accuracy given the same amount of local measurement data (see Supplementary Note 7). 

To highlight the importance of our representation neural network for good prediction accuracy, we replaced it with principal component analysis (PCA) and trained individual prediction neural networks with PCA-generated representations as input. This simplification resulted in a complete failure to predict any (B1) or (B2) properties (see Supplementary Note 5).  This reveals that PCA cannot extract the essential information of quantum states from their limited measurement data, a task successfully accomplished by our trained representation network.


\begin{figure*}
    \includegraphics[width=0.6\textwidth]{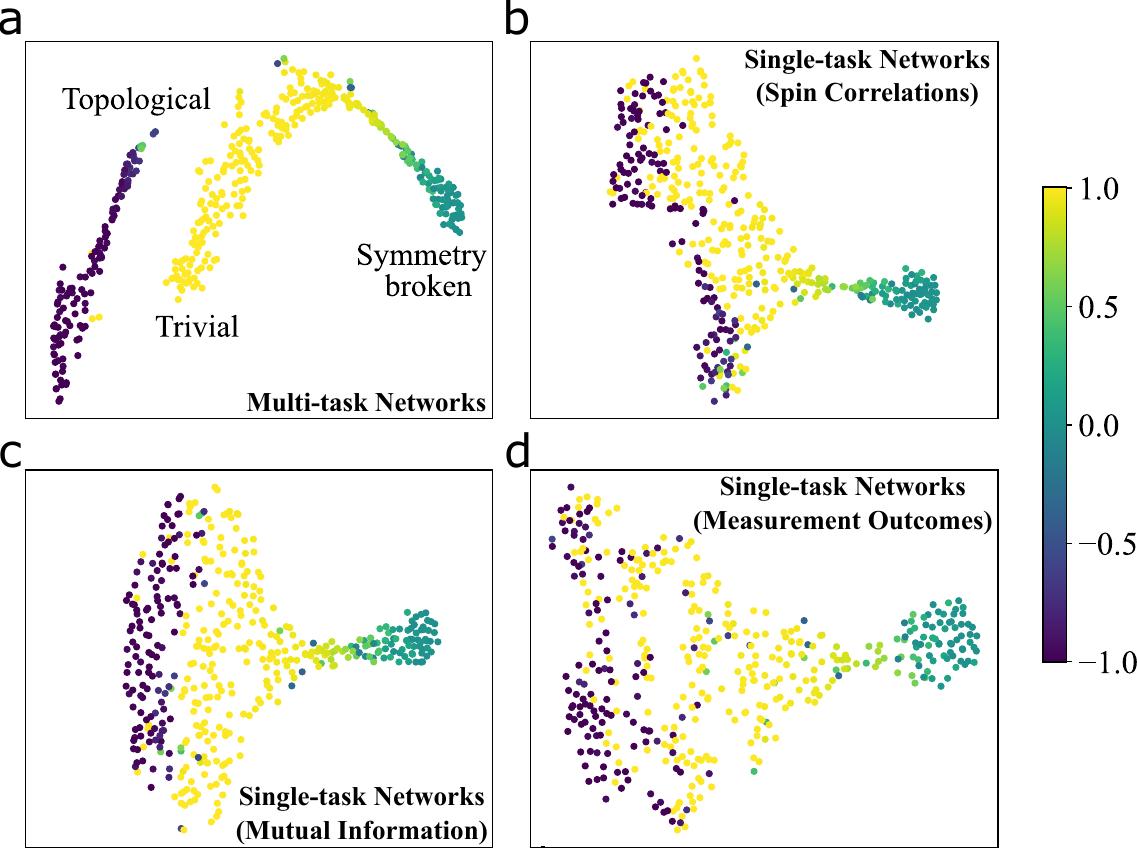}
    \caption{2D projections of the state representations for bond-alternating XXZ model obtained with the t-SNE algorithm. The color of  each data point indicates the true value of many-body topological invariant $\mathcal Z_{\text{R}}$ of the corresponding ground state. Subfigure {\bf a} corresponds to the state representations produced for predicting both spin correlations and mutual information. Whereas, Subfigures {\bf b, c} and {\bf d} correspond to the state representations produced for predicting spin correlations, mutual information and measurement outcome distributions respectively. }
    \label{fig7}
\end{figure*}

\begin{figure*}
    \centering
    \includegraphics[width=0.6\textwidth]{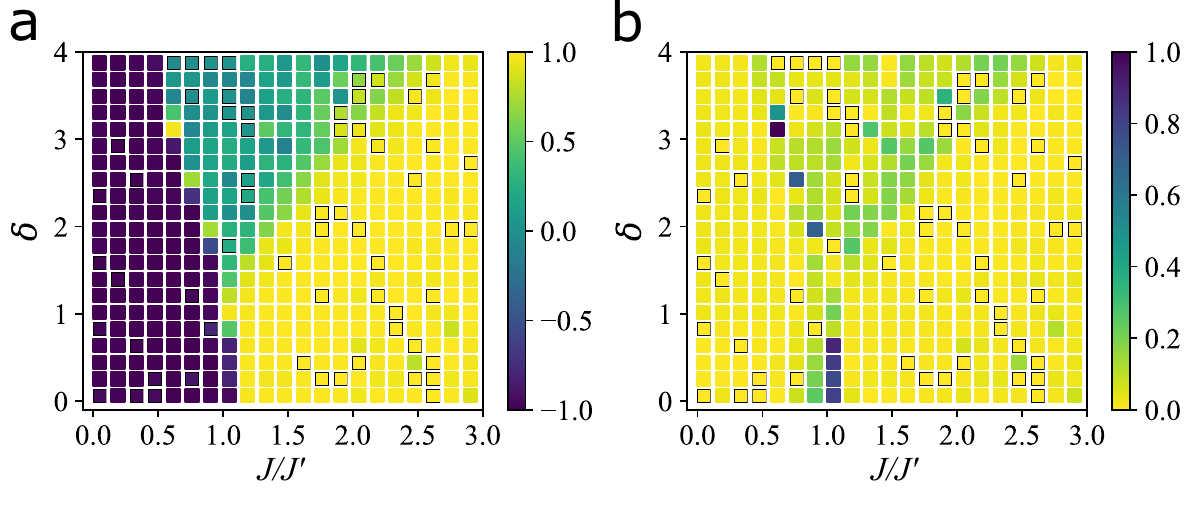}
    \caption{Predictions of many-body topological invariants.   Subfigure {\bf a} shows the predictions of $\mathcal Z_{\text{R}}$ for the ground states corresponding to all pairs of parameters $(J/J', \delta)$ together with the true values of $60$ reference states marked by grey squares. Subfigure {\bf b} shows the absolute values of the differences between predictions and the ground truths. }
    \label{fig8}
\end{figure*}

We show that, even in the larger-scale example considered in this section,   the state representations obtained through multi-task training contain  information about the quantum phases of matter. 
In Figure~\ref{fig7}{\bf a},
we show the  2D-projection of the state representations.
The data points corresponding to ground states in the topological SPT phase,  the trivial SPT phase and the symmetry broken phase appear to  be  clearly separated into three clusters,  while the latter two connected by a few data points corresponding to  ground states across the phase boundary.  
A few points, corresponding  to  ground states near phase boundaries of the topological SPT phase,   are incorrectly  clustered by the t-SNE algorithm.  
The origin of the problem  is  that the correlation length of  ground states near phase boundary becomes longer, and therefore the measurement statistics on nearest-neighbour-three qubit subsystems cannot capture sufficient information for predicting the correct  phase of matter.

We further examine if the single-task neural networks above can correctly classify the three different phases of matter. 
We project the state representations produced by each single-task neural network onto 2D planes by the t-SNE algorithm, as shown by Figures~\ref{fig7}{\bf b} and \ref{fig7}{\bf c}. The pattern of projected representations in Figure~\ref{fig7}{\bf b} implies that when trained only with the values of spin correlations, the neural network cannot distinguish the topological SPT phase from the trivial SPT phase.
The pattern in Figure~\ref{fig7}{\bf c} indicates that when trained solely with mutual information, the performance of clustering is slightly improved, but still cannot explicitly classify these two SPT phases.
We also project the state representations produced by the neural network for predicting measurement outcome statistics~\cite{zhu2022} onto a 2D plane.  The resulting pattern, shown in Figure~\ref{fig7}{\bf d}, shows  that the topological SPT phase and the trivial SPT phase cannot be correctly classified either. These observations indicate that  a multi-task approach, including both the properties of mutual information and spin correlations,  
is necessary to capture the difference between the topological SPT phase and the trivial SPT phase.


The emergence of clusters related to different phases of
matter suggests that the state representation produced
by our network also contains quantitative information
about the topological invariant $\mathcal Z_{\text{R}}$.  
To extract this information, we use an additional neural network, which maps the state representation into a prediction of  $\mathcal Z_{\text{R}}$, We train this additional network by randomly
selecting $60$ reference states (marked by grey squares in
Figure~\ref{fig8}) out of the set of $441$ fiducial states, and by
minimizing the prediction error on the reference states.
The predictions together with 60 exact values of the reference states are shown in Figure~\ref{fig8}{\bf a} 
The absolute values of the differences between the predictions
and ground truths are shown in Figure~\ref{fig8}{\bf b}.   
The predictions are close to the ground truths, except for the ground states near the phase boundaries, especially the boundary of topological SPT phase. The mismatch at the phase boundaries corresponds the state representations incorrectly clustered in Figure~\ref{fig7}{\bf a}, suggesting our network struggles to learn long-range correlations at phase boundaries.

\begin{figure*}
\centering
    \includegraphics[width=0.7\textwidth]{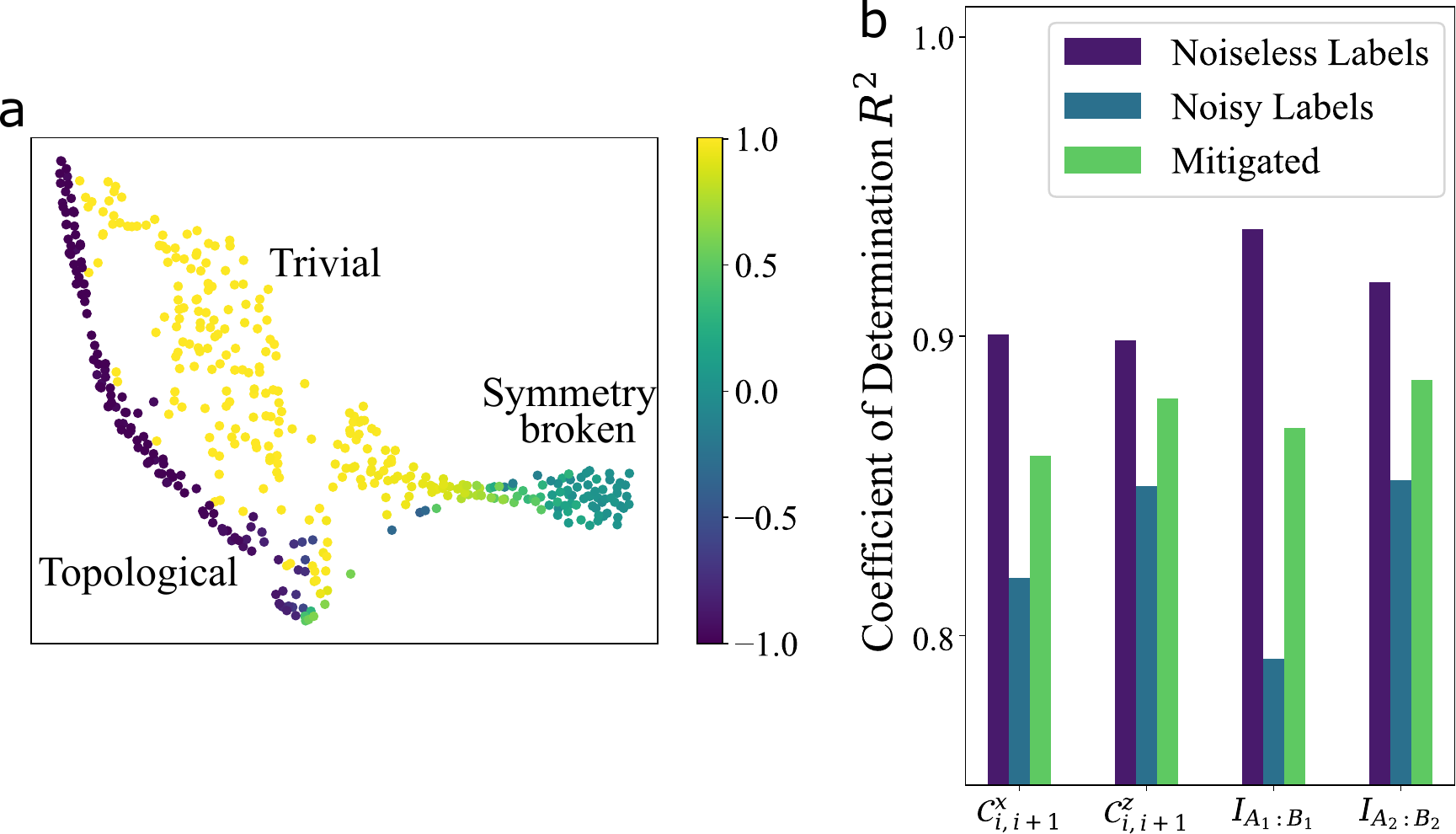}
     \caption{Predictions of properties of $50$-qubit systems made by a neural network trained over the data of $10$-qubit systems. Subfigure {\bf a} shows the 2D projections of state representations via t-SNE algorithm. Subfigure {\bf b} shows the coefficient of determination for the predictions of properties using noiseless training labels and noisy training labels, as well as the predictions after error mitigation.}
     \label{fig9}
\end{figure*}

\subsection*{Generalization to quantum systems of larger size}
We now show that our model is capable of extracting features 
that are transferable across different system sizes. 
To this purpose,  
we use a training dataset generated from $10$-qubit ground states of the bond-alternating XXZ model~(\ref{eq:XXZmodel}) and then we use the trained network  
to generate state representations from the local measurement data of each $50$-qubit ground state of~(\ref{eq:XXZmodel}).

Figure~\ref{fig9}{\bf a} shows that inputting the state representations into  the t-SNE algorithm still gives rise to  clusters according to the three distinct phases of matter.
This observation suggests that the neural network can effectively classify different phases of the bond-alternating XXZ model, irrespective of the system size. In addition to clustering larger quantum states, the representation network also facilitates the prediction of quantum properties in the larger system. To demonstrate this capability, we employ $40$ reference ground states of the $50$-qubit bond-alternating XXZ model, which are only half size of the training dataset used for $10$-qubit system, to train two prediction networks: one for spin correlations and the other for mutual information. Figure~\ref{fig9}{\bf b} shows the coefficients of determination for each prediction, which exhibit  values around 0.9 or above.  Figure~\ref{fig9}{\bf b} also shows the impact of inaccurate labelling of the ground states on our model.  
  In the reported experiments,  we assumed that $10\%$ of the labels in the training dataset corresponding to $40$ reference states are randomly incorrect, while the remaining $90\%$ are accurate. Without any mitigation, we observe that the errors substantially impacts the accuracy of our predictions.   On the other hand,  employing a technique of noise mitigation during the training of prediction networks (see Supplementary Note 7) can  effectively reduce the impact of the incorrect labels.

\section{Discussion}
\label{sec:discussion}
The use of short-range local  measurements is a  key distinction between our work and prior approaches approaches using randomized measurements~\cite{huang2020,huang2022provably,elben2020,elben2023randomized}.  
 Rather than measuring all spins together, we employ randomized Pauli measurements on small groups of neighboring sites. This feature is appealing for practical applications, as measuring correlations among large numbers of sites is generally challenging.   In Supplementary Note 5, we show that  classical shadow estimation cannot be directly adapted to  the scenario where only short-range local measurements are available.  On the other hand, the restriction to short-range local measurements implies that the applicability of our method is limited to 
many-body quantum states with a constant correlation length, such as the ground state within an SPT phase. 



A crucial aspect of our neural network model is its ability to  generate 
a latent state representation that integrates different pieces of information, corresponding to multiple physical properties.  
Remarkably, the state representations appear to capture  information about properties  beyond those encountered in training.  This feature allows for unsupervised classification of phases of matter, applicable not only to in-distribution Hamiltonian ground states but also to out-of-distribution quantum states, like those produced by random circuits. 
The model also appears to be able to generalize from smaller to larger quantum systems, which makes it an effective tool for  exploring intermediate-scale quantum systems.

For new quantum systems, whose true phase diagrams is still unknown, 
discovering phase diagrams in an unsupervised manner is a major challenge.
This challenge can potentially be addressed  by combining our neural network with consistency-checking, similar to the approach in Ref.~\cite{van2017}. The idea is to start with an initial, potentially inaccurate, phase diagram ansatz constructed from limited prior knowledge, for instance, the results of clustering. Then, one can randomly select a set of reference states, labeling them according to the ansatz phases. Based on these labels,  a separate neural network is trained to predict phases.  
Finally, the ansatz can be revised based on the deviation with the network's prediction, and the procedure can be iterated until it converges to a stable ansatz. In Supplementary Note 8, we provide examples of this approach,  
leaving the development of a full algorithm for autonomous discovery of phase diagram as  future work.

\section{Methods}
\label{sec:methods}
{\em Data generation.} Here we illustrate the procedures for generating training and test datasets. For the one-dimensional cluster-Ising model 
, we obtain measurement statistics and values for various properties in both the training and test datasets through direct calculations, leveraging the ground states solved by exact algorithms. In the case of the one-dimensional bond-alternating XXZ model, we first obtain approximate ground states represented by matrix product states~\cite{fannes1992finitely,perez2007matrix} using the density-matrix renormalization group (DMRG)~\cite{schollwock2005density} algorithm. Subsequently, we compute the measurement statistics and properties by contracting the tensor networks. For the noisy measurement statistics because of finite sampling, we generate them by sampling from the actual probability distribution of measurement outcomes. More details are provided in Supplementary Note 1. 

{\em Representation Network.}
The representation network operates on pairs of measurement outcome distributions and the parameterization of their corresponding measurements, denoted as ${(\bm{d}_i,\bm{m}_i)}_{i=1}^m$ associated with a state $\rho$. This network primarily consists of three multilayer perceptrons (MLPs)~\cite{gardner1998artificial}. The first MLP comprises a four-layer architecture that transforms the measurement outcome distribution into $\bm{h}^d_i$, whereas the second two-layer MLP maps the corresponding $\bm{m}_i$ to $\bm{h}^m_i$:

\begin{align*}\label{eq:rep1}
   \bm{h}^d_i &= {\rm MLP}_1(\bm{d}_i),\\
   \bm{h}^m_i &= {\rm MLP}_2(\bm{m}_i).\\
\end{align*}

Next, we merge $\bm{h}^d_i$ and $\bm{h}^m_i$, feeding them into another three-layer MLP to obtain a partial representation denoted as $\bm{r}_i$ for the state:

\begin{equation}
    \bm{r}_\rho^{(i)} =  {\rm MLP}_3([\bm{h}^d_i, \bm{h}^m_i]).
\end{equation}

Following this, we aggregate all the $\bm{r}_i$ representations through an average pooling layer to produce the complete state representation, denoted as $\bm{r}_\rho$:

\begin{equation}
     \bm{r}_\rho = \frac{1}{s} \sum_{i=1}^s \bm{r}_i.
\end{equation}

Alternatively, we can leverage a recurrent neural network equipped with gated recurrent units (GRUs)~\cite{chung2014empirical} to derive the comprehensive state representation from the set $\{\bm{r}_i\}_{i=1}^m$:

\begin{align*}
   \bm{z}_i &= {\rm sigmoid}(W_z\bm{r}_\rho^{(i)} + U_z \bm{r}_\rho^{(i-1)} + \bm{b_z})  ,\\
   \hat{\bm{h}}_i &= {\rm tanh}(W_h \bm{r}_\rho^{(i)} + U_h( \bm{z}_i  \odot \bm{h}_{i-1}) + \bm{b_h}),\\
   \bm{h}_i &= (1- \bm{z}_i) \odot \bm{h}_{i-1} + \bm{z}_i \odot \hat{\bm{h}}_i,\\
   \bm{r}_\rho &= \bm{h}_{m},
\end{align*}
where $W,U,\bm{b}$ are trainable matrices and vectors. The architecture of the recurrent neural network offers a more flexible approach to generate the complete state representation; however, in our experiments, we did not observe significant advantages compared to the average pooling layer.

\begin{figure*}
\centering
    \includegraphics[width=0.85\textwidth]{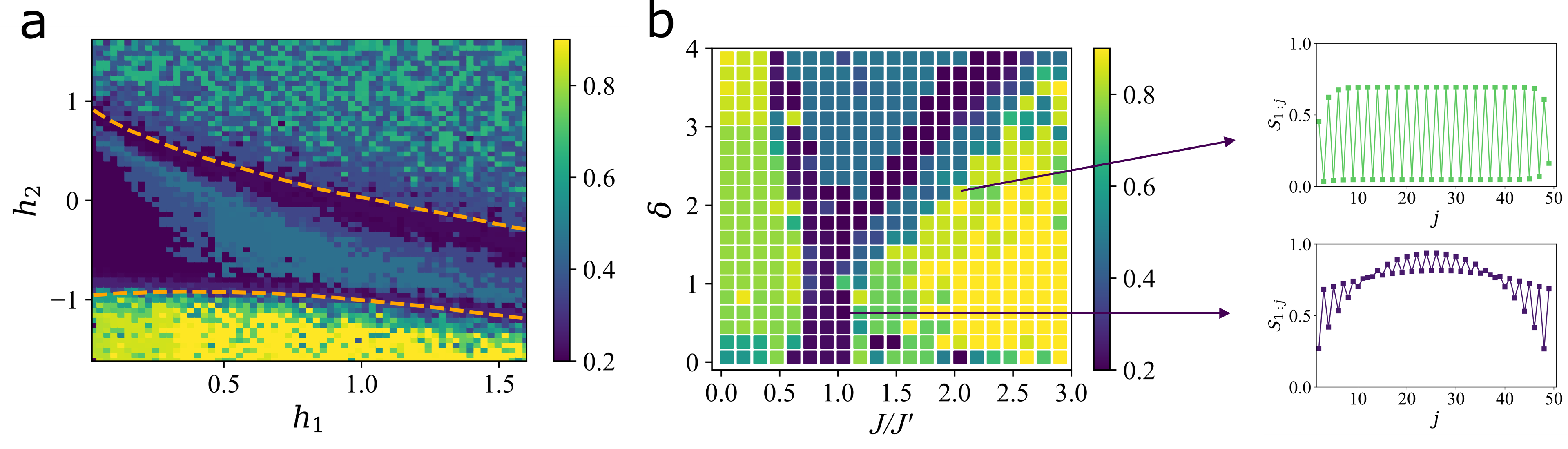}
     \caption{ A measure of reliability of state representations. Subfigure {\bf a} shows the reliability of the representation of each ground state of the cluster-Ising model. Subfigure {\bf b} illustrates the reliability of the representation of each ground state of the bond-alternating XXZ model, together with the true subsystem entanglement entropy for two example states: one is near the phase boundary and the other is further away from the phase boundary.}
     \label{fig10}
\end{figure*}

{\em Reliability of Representations.} 
The neural network can assess the reliability of each state representation by conducting contrastive analysis within the representation space.  
Figure~\ref{fig10} shows a measure of the reliability of each state representation, which falls in the region $[0, 1]$, for both the cluster-Ising model and the bond-alternating XXZ model. As this measure increases from $0$ to $1$, the reliability of the corresponding prediction strengthens, with values closer to $0$ indicating low reliability and values closer to $1$ indicating high reliability.
Figure~\ref{fig10}{\bf a} indicates that the neural network exhibits lower confidence for the ground states in SPT phase than those in the other two phases, with the lowest confidence occurring near the phase boundaries.
Figure~\ref{fig10}{\bf b} shows that the reliability of predictions for the ground states of the XXZ model in two SPT phases are higher than those in the symmetry broken phase, which is due to the imbalance of training data, and that the predictions for quantum states near the phase boundaries have the lowest reliability.   
Here, the reliability is associated with the  distance between the state representation and its cluster center in the representation space. We adopt this definition based on the intuition that for a quantum state that the model should exhibit higher confidence for quantum states that cluster more easily. 

Distance-based methods~\cite{lee2018simple, sun2022out} have proven effective in the task of Out-of-Distribution detection in classical machine learning. This task focuses on identifying instances that significantly deviate from the data distribution observed during training, thereby potentially compromising the reliability of the trained neural network.  
Motivated by this line of research, we present a contrastive methodology for assessing the reliability of representations produced by the proposed neural model. Denote the set of representations corresponding quantum states as $\{\bm{r}_{\rho_1},\bm{r}_{\rho_2},\cdots,\bm{r}_{\rho_n}\}$. We leverage reachability distances, $\{d_{\rho_j}\}_{j=1}^n$, derived from the OPTICS (Ordering Points To Identify the Clustering Structure) clustering algorithm~\cite{ankerst1999optics} to evaluate the reliability of representations, denoted as $\{rv_{\rho_j}\}_{j=1}^n$:

\begin{align*}
   \{d_{\rho_j}\}_{j=1}^n &= {\rm OPTICS}( \{\phi(\bm{r}_{\rho_j})\}_{j=1}^n), \\
   rv_{\rho_j} &= \frac{\exp(-d_{\rho_k})}{\max_{k=1}^n \exp(-d_{\rho_k})},
\end{align*}

where $\phi$ is a feature encoder. In the OPTICS clustering algorithm, a smaller reachability distance indicates that the associated point lies closer to the center of its corresponding cluster, thereby facilitating its clustering process. Intuitively, a higher density within a specific region of the representation space indicates that the trained neural model has had more opportunities to gather information from that area, thus enhancing its reliability.  Our proposed method is supported by similar concepts introduced in ~\cite{sun2022out}. More details are provided in Supplementary Note 3.

{\em Prediction Network.} For each type of property associated with the state, we employ a dedicated prediction network responsible for making predictions. Each prediction network is composed of three MLPs. The first MLP takes the 
 state representation $\bm{r}_\rho$ as input and transforms it into a feature vector $\bm{h}^{\bm{r}}$ while the second takes the query task index $q$ as input and transforms it into a feature vector $\bm{h}^q$. The second MLP operates on the combined feature vectors $[\bm{h}^{\bm{r}}, \bm{h}^q]$ to produce the prediction $f_{q}(\rho)$ for the property under consideration:

\begin{align*}\label{eq:rep1}
    \bm{h}^{\bm{r}} &= {\rm MLP}_4(\bm{r}_\rho), \\
   \bm{h}^q &= {\rm MLP}_5(q),\\
  f_{q}(\rho) &=  {\rm MLP}_6([\bm{h}^{\bm{r}},\bm{h}^q ]).\\
\end{align*}

{\em  Network training.} We employ the stochastic gradient descent~\cite{bottou2012stochastic} optimization algorithm and the Adam optimizer~\cite{kingma2014adam} to train our neural network. In our training procedure, for each state within the training dataset, we jointly train both the representation network and the prediction networks associated with one or two types of properties available for that specific state. This training is achieved by minimizing the difference between the predicted values generated by the network and the ground-truth values, thus refining the model's ability to capture and reproduce the desired property characteristics. The detailed pseudocode for the training process can be found in Supplementary Note 2.

{\em Hardware.}
We employ the PyTorch framework \cite{paszke2019pytorch} to construct the multi-task neural networks in all our experiments and train them with two NVIDIA GeForce GTX 1080 Ti GPUs.

{\em Acknowledgements.}
We thank  Ge Bai, Dong-Sheng Wang, Shuo Yang and Yuchen Guo  for the helpful discussions on many-body quantum systems.
This work was supported by funding from the Hong Kong Research Grant Council through grants no.\ 17300918 and no.\ 17307520, through the Senior Research Fellowship Scheme SRFS2021-7S02, and the John Templeton Foundation through grant  62312, The Quantum Information Structure of Spacetime (qiss.fr).  YXW acknowledges funding from the National Natural Science Foundation of China through grants no.\ 61872318. Research at the Perimeter Institute is supported by the Government of Canada through the Department of Innovation, Science and Economic Development Canada and by the Province of Ontario through the Ministry of Research, Innovation and Science. The opinions expressed in this publication are those of the authors and do not necessarily reflect the views of the John Templeton Foundation.

\bibliography{refs}

\end{document}